# Global total precipitable water variations and trends during 1958-2021

Nenghan Wan[1], Xiaomao Lin[1*], Roger A. Pielke Sr.[2], Xubin Zeng[3], and Amanda M. Nelson[4]

[1]Department of Agronomy, Kansas Climate Center, Kansas State University, Manhattan, KS.
[2]Cooperative Institute for Research in Environmental Sciences, University of Colorado, Boulder, CO.
[3]Climate Dynamics and Hydrometeorology Center at the University of Arizona, Tucson, AZ.
[4]National Center of Alluvial Aquifer Research, USDA-ARS Sustainable Water Management Research Unit, Stoneville, MS.

*Correspondence to*: Xiaomao Lin (xlin@ksu.edu)

**Abstract.** Global responses of the hydrological cycle to climate change have been widely studied but uncertainties of temperature responses to lower-tropospheric water vapor still remain. Here, we investigate the trends in global total precipitable water (TPW) and surface temperature from 1958 to 2021 using improved ERA5 and JRA-55 reanalysis datasets and further validate these trends by using radiosonde, Atmospheric Infrared Sounder (AIRS), and Microwave Satellite (SSMI(S)) observations. Our results indicate a global increase in total precipitable water (TPW) of 0.66% per decade according to ERA5 data and 0.88% per decade in JRA-55 data. These variations in TPW reflect the interactions of global warming feedback mechanisms across different spatial scales. Our results also revealed a significant near-surface temperature ($T_{2m}$) warming trend at the rate of 0.14 K dec$^{-1}$ and a strong water vapor response to temperature at a rate of 4-6% K$^{-1}$ globally, with land areas warming approximately twice as fast as the oceans. The relationship between TPW and $T_{2m}$ or surface skin temperature ($T_s$) showed a variation around 6 - 8% K$^{-1}$ in the 15-60 °N latitude band, aligning with theoretical estimates from the Clausius–Clapeyron equation.

## 1 Introduction

As the most critical greenhouse gas in the Earth's atmosphere, water vapor plays a key role in atmospheric processes from the microscale (including the formation of clouds and precipitation) to the global scale, and is related to the Earth's radiation budget, hydrological cycle, and climate change (Held and Soden, 2006; Lacis et al., 2010; Ruckstuhl et al., 2007). The amount of water vapor is primarily controlled by the air temperature when the relative humidity (especially over the ocean) remains unchanged in the low troposphere. The total precipitable water (TPW), also known as the column-integrated amount of water vapor from the surface to the top of the atmosphere, increases by 6-7% with a 1 K increase in air temperature according to the Clausius-Clapeyron equation and thus enhances the strength of global warming with strong positive feedback due to the greenhouse effect (Held and Soden, 2006). Therefore, evaluating the long-term trend of TPW change and its relationship with temperature is important for understanding the role of water vapor in climate change and the impact of water vapor feedback on global warming.

Because of the short residence time of water vapor in the atmosphere, studies in terms of long-term water vapor trends and their variability still face challenges. The two main categories of water vapor data include observations collected from various weather station networks and satellites, as well as reanalysis datasets integrated with simulations and observations. The former has difficulty in evaluating long-term trends because of sparsely distributed stations, with discontinuous and inhomogeneous observations (Dee et al., 2011), while in some of the previous reanalysis datasets, biases and errors in observations that are assimilated into the reanalyses system may affect the reanalysis dataset's quality and therefore cause concerns about their reliability for detecting climate trends (Dai et al., 2011; Schröder et al., 2016; Trenberth et al., 2011). Numerous studies have analysed trends and variations in atmospheric water vapor distribution on both global and regional scales, primarily utilizing early versions of reanalysis datasets and satellite observations, albeit over relatively short study periods (e.g., Borger et al., 2022; Parracho et al., 2018; Wang et al., 2016; Zhang et al., 2019, 2021). New generation reanalysis datasets (JRA-55 and ERA5) provide a better option for climate studies because of advanced modelling and data assimilation systems, with better accuracy, and fewer homogeneity issues (Hersbach et al., 2020; Kobayashi et al., 2015). JRA-55 showed a better performance in studying multidecadal variability and climate change than previous reanalysis datasets (Kobayashi et al., 2015) and is reliable even for the pre-satellite era (Kikuchi, 2020). Many studies have confirmed that ERA5 was the best or among the highest-performing reanalysis products (Taszarek et al., 2021; Yuan et al., 2023). Therefore, the latest ERA5 and JRA-55 were selected in this study to analyze long-term TPW changes and their relationship with temperatures at regional and global scales.

The near-surface air temperature (2m air temperature, $T_{2m}$) describes the thermodynamic temperature at a 1.5-2 m height while surface skin temperature ($T_s$) refers to 'radiometric surface temperature' that is governed by the terrestrial radiation balance (Jin et al., 1997; Jin and Dickinson, 2010). Recent studies investigated how the TPW responds to changes in surface temperature using modelling estimation, satellite observations, or ground-based observation systems (Allan et al., 2022; Alshawaf et al., 2017; Borger et al., 2022; O'Gorman and Muller, 2010; Wang et al., 2016; Yuan et al., 2021), and many global surface temperature observational datasets use sea surface temperature (SST) over ocean and $T_{2m}$ over land (e.g., HadCRUT, Morice et al., 2021). Different from prior studies, we use the newest and longest available reanalyses datasets to analyze the relationship between trends in TPW and surface temperature ($T_{2m}$ and $T_s$) over the past 64 years, discuss the difference of $T_{2m}$ and $T_s$, and the potential discontinuities at global-scale, and compare results with satellite and radiosonde data.

In this study, we focus on answering the following questions: (1) To what extent has the TPW changed from 1958 to 2021 and what is the difference between the variations of $T_{2m}$ and $T_s$ on a multi-decade scale? (2) What is the relationship between trends of TPW changes and temperature changes? Specifically, we analysis the trends in TPW and two temperatures (i.e, $T_{2m}$ and $T_s$), as well as their relationships. We then examine with satellite and radiosonde data. Lastly, we discuss the difference of results among datasets and their potential discontinuities in datasets. Therefore, this paper is organized as follows. In section 2, we introduce the datasets and methods. In section 3, the TPW variations and the differences of $T_{2m}$ and $T_s$ are compared with radiosonde and satellite observations over land and ocean. Sections 4 and 5 provide the discussion and conclusion.

## 2 Data and Methods

### 2.1 Datasets

Two reanalysis datasets containing TPW and temperature ($T_{2m}$, $T_s$) from 1958 to 2021 were used in this study, these being ERA5 (Hersbach et al., 2020) and JRA-55 (Kobayashi et al., 2015). For ERA5, skin temperature, $T_s$, is the temperature estimated from the surface energy balance. For JRA-55, skin temperature is a diagnostic variable computed from the surface upward longwave radiation under the assumption that the surface is a black body. We used monthly reanalysis datasets from January 1958 to December 2021 for TPW and temperatures. Both TPW and temperature variables in these two datasets were regirded into 1º x 1º resolution data using a bilinear interpolation method before analyzing (Zhuang, 2018).

The Atmospheric Infrared Sounder (AIRS) instrument captures a precisely calibrated, spectrally detailed dataset of both infrared and microwave radiances. It offers quality temperature and humidity profiles throughout the troposphere (Tian et al., 2020). Satellite observation from version 7 of AIRS (Tian et al., 2020) was used for comparison with the reanalysis from 2003 to 2021. The total column water vapor (kg m$^{-2}$) is calculated as the average of daytime (TotH2OVap_A) and nighttime modes (TotH2OVap_D). In addition, the column-integrated water vapor from the Special Sensor Microwave Imager and the Special Sensor Microwave Imager Sounder (SSMI(S)) from 2003-2021 (Wentz et al., 2015) are selected for comparison with reanalysis datasets.

We also used *in-situ* observations from the Radiosounding HARMonization (RHARM) dataset (Madonna et al., 2022), which is a completely independent data source of reanalysis datasets. RHARM provides homogenized temperature and relative humidity profiles at two observation times (0000 and 1200 UTC) for radiosonde stations globally from 1979 to 2019. Therefore, $T_{2m}$ used in this study from radiosonde observation was taken from the temperature observed at the lowest layer. Relative humidity and temperature from the surface to 500 hPa level are used to calculate precipitable water.

### 2.2 Methods

Monthly TPW and temperature anomalies were calculated by the base period of 1958-2021 for reanalysis datasets. We divided the globe into tropical regions (23.5°S–23.5°N), temperate (23.5°S–66.5°S in Southern Hemisphere (SH) and 23.5°N–66.5°N in Northern Hemisphere (NH)), and polar regions (66.5°S–90°S in SH and 66.5°N–90°N in NH). The regional average values, presented by different latitude bins, were calculated with a cosine (latitude) weighting factor to account for the convergence of grid points for each region. For the global distribution, all datasets were resampled into 1º x 1º resolution using spatial-averaging resampling, then the area-weighted average of anomalies was computed to formulate a global time series.

For all trend analyses in TPW and temperature ($T_s$, $T_{2m}$) series, we selected the Seasonal Kendall (SK) test (Hirsch et al., 1982) using the Theil-Sen slopes (Sen 1968; Theil, 1992) to calculate relative TPW trends (% dec$^{-1}$) (Zhai and Eskridge, 1997) and temperature trends for both $T_s$ and $T_{2m}$ (K dec$^{-1}$). The statistical significance of all linear correlations and trends used was performed at a 95% confidence level for all analyses conducted in this study. To obtain the percent change of TPW with respect to temperature, first the long-term trend of TPW (% dec$^{-1}$) and temperature trends (K dec$^{-1}$) are calculated, then the ratio of TWP trends and temperature trends is calculated for the TPW responses to temperature.

Due to daily data missing in radiosonde observations, valid data required at least 10 days of data available within a month, and at least two-thirds of the total months had to have valid monthly data (345 months for 1979-2019). Each month had to have at least 28 years of valid data at each station to calculate monthly climatology. Trend analysis is also from Sen's estimate of the slope (Sen 1968) for these observations. Thereby, a set of 331 radiosonde stations were selected in this study. To facilitate the comparison, interpolated reanalysis data from radiosonde stations were selected to compare with the observation data. We calculated the trend difference between reanalyses and observations to assess how effectively each reanalysis product captures long-term changes in observed TPW over 1979-2019.

To detect potential discontinuity within temporal sequence of total water vapor and temperature in observation data, we applied penalized maximal F test (PMF test, Wang et al., 2006; Wang, 2008) to detect temporal discontinuity points on averaged monthly anomaly over ocean, land, NH polar, North America, and Tropical land regions during 1958-2021. The discontinuities were documented if testing statistic's p-value was less than 0.05. The magnitudes of shifts at discontinuity points, also called step sizes, are calculated for each discontinuity point. Then, discontinuity point positions were compared with changes in observing measurement systems. We defined the temporal coincidence when the discontinuity point positions were within a 3-month window along with observing measurement systems. Lastly, after identifications of discontinuity point's magnitudes and positions, the times series were adjusted using the quantile-matching (QM) procedure (Wang et al., 2010). Trend comparisons were conducted before and after QM-adjusted, and the RHtestsV4 tool is used for these calculations (Wang and Feng, 2013).

## 3 Results and Discussion

### 3.1 TPW Trends

Figure 1 shows the decadal trends in TPW calculated from ERA5 and JRA-55 between 1958 and 2021. In general, the two reanalysis datasets showed similar trend patterns of global TPW distribution, with upward trends as the dominant change of TPW, indicating a rise in moisture in response to global warming since the twentieth century (Santer et al., 2006). TPW trends are largely positive and statistically significant over North America, tropical land regions (± 23.5° band), and the NH polar region. On the other hand, decreasing TPW trends show a dipole structure over the Southern Ocean in both reanalysis datasets which can be attributed to the change of the ENSO phase (Trenberth et al., 2005). In contrast, these two reanalysis datasets show opposite trends over eastern Africa including the Sudan and Ethiopia where ERA5 trends are positive, whereas JRA-55 trends are negative. The reason for opposite trends in reanalyses is likely due to the different representations of large-scale moisture transport, surface-atmosphere processes, and their data assimilation (Parracho et al., 2018). Chen and Liu (2016) and Parracho et al. (2018) also showed decreasing trends in North Africa using ERA-Interim reanalysis for 1979-2014 and 1980-2016, and the biases in rainfall over West Africa might also exist in ERA-Interim reanalysis (Dunning et al., 2016).

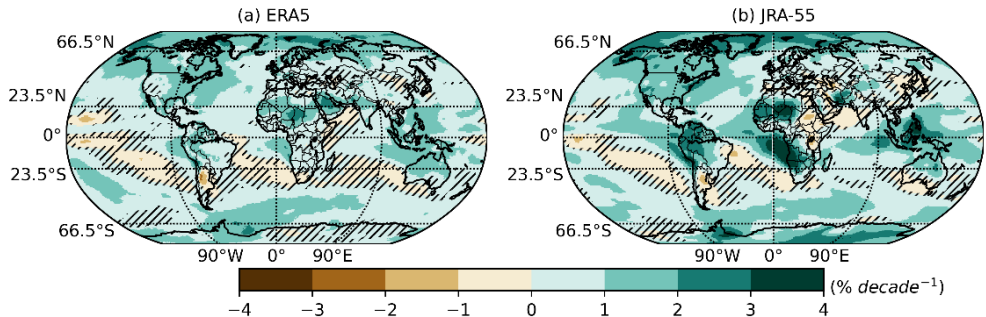

**Figure 1.** Monthly total precipitable water (TPW) trend (% dec$^{-1}$) from 1958 to 2021 from **(a)** ERA5 and **(b)** JRA-55. The hatch areas represent trends that are not statistically significant at a 95% confidence level.

Figs. 2a-e show the temporal variations of monthly TPW anomalies over tropical, temperate, and polar regions. Overall, in terms of latitude bands, these trends differ little between the two reanalysis datasets. Specifically, ERA5 gives a 0.61 % dec$^{-1}$ moistening rate whereas the JRA-55 shows a 0.97 % dec$^{-1}$ moistening rate over the tropical regions. A notable decrease in TPW is shown during the 1980s to1990s over tropical regions in ERA5, which has been noted before and is consistent with satellite microwave data (Allan et al., 2022). Trends range from 0.82 to 1.95 % dec$^{-1}$ over temperate and polar regions in NH and are about 0.44 % dec$^{-1}$ over temperate and 0.84 % dec$^{-1}$ over polar regions in SH for both reanalysis datasets. The TPW trends increase in the NH is larger than those in the SH although they showed strong month-to-month variabilities (coefficient of variation, CV, is 0.7 on average) in the NH, which are consistent with the global distribution (Fig. 1).

Figures 2f-h depict these trends of TPW estimated over the ocean, land, and the globe. It is noticeable that monthly TPW anomalies showed a consistent and statistically significant increase, yielding a rate of 0.57, 0.9, and 0.66 % dec$^{-1}$ for ERA5 and 0.84, 1.05, and 0.88 % dec$^{-1}$ for JRA-55 over ocean, land, and the globe, respectively. The monthly TPW anomalies showed a similar variability over ocean, land, and globally (CVs, around 0.4), and displayed significant interannual variations, which had been dominated by ENSO events (Trenberth et al., 2005; Wang et al., 2016). The distinct turning points may be attributed to the intensity of ENSO events. For example, one of the most powerful ENSO events during the 1997/1998 El Niña led to a significant tropical TPW increase, attributable to the warming in the equatorial Pacific (Wagner et al. 2005). The 2015/2016 El Niño caused a moistening TPW trend over tropical regions (Garfinkel et al., 2018).

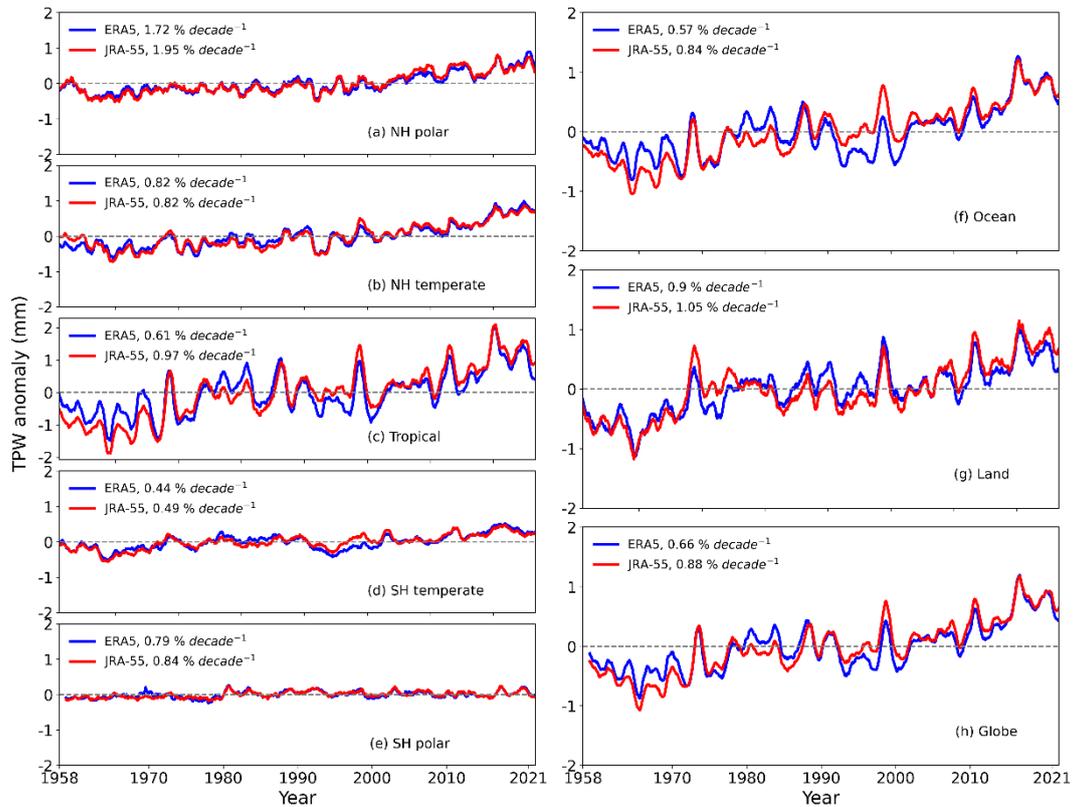

**Figure 2.** Time series of monthly TPW anomalies (mm) over **(a)** the Northern Hemisphere (NH) polar, **(b)** NH temperate, **(c)** Tropical, **(d)** Southern Hemisphere (**SH**) temperate, **(e)** SH polar, **(f)** ocean area, **(g)** land area, and **(h)** the globe from the ERA5 (blue curves) and JRA-55 (red curves) during the period 1958–2021. A 12-month running smoother was applied to each time series shown. All trends, shown in units of % per decade, were statistically significant at a 95% confidence level.

**3.2 Trends in temperature**

The global trends of temperature ($T_{2m}$ and $T_s$) between 1958-2021 exhibit widespread warming over large regions in both reanalysis datasets, except for some of small areas located in the SH and northern Atlantic (Fig. 3). Large land areas experience warming temperature changes of greater magnitude compared to the surrounding oceans (Fig. 3). Both the trends of $T_{2m}$ and $T_s$ show a remarkably similar distribution, with a clear warming trend observed in the NH. The strongest warming occurred in the Arctic (Fig. 3). Greater warming is observed over mid-latitude than tropical regions in NH, consistent with Zeng et al. (2021). Cooling in the North Atlantic is also observed by Li et al. (2022) when using observation datasets. This arises from strengthened local convection and heat dissipation from the ocean induced by the overlying atmosphere. For the SH, the polar region also experiences the strongest warming trends, similar to the trend's magnitudes by Clem et al. (2020). There is a slight difference between two reanalysis datasets for differential trends (ΔT) between $T_s$ and $T_{2m}$ at the global scale, especially over the ocean area where ΔT trends are negative in ERA5 and positive in JRA-55. The ΔT shows a similar pattern only over the tropic land region (Fig. 3f). Although air and surface temperatures are closely related, they are physically distinct. The difference may be affected by changes of regional and seasonal vegetation ecosystems, land use, and land cover (Gulev et al., 2021; Masson-Delmotte, 2022). The ability of atmospheric reanalyses to effectively constrain variations in $T_s$ and $T_{2m}$ trends is limited because the sea surface temperatures in atmospheric reanalyses were obtained from globally observed and interpolated products (Rayner et al. 2003). In contrast, the estimation of air temperature depends on model parameterizations and assimilated observations, which do not incorporate marine air temperature (Simmons et al., 2017).

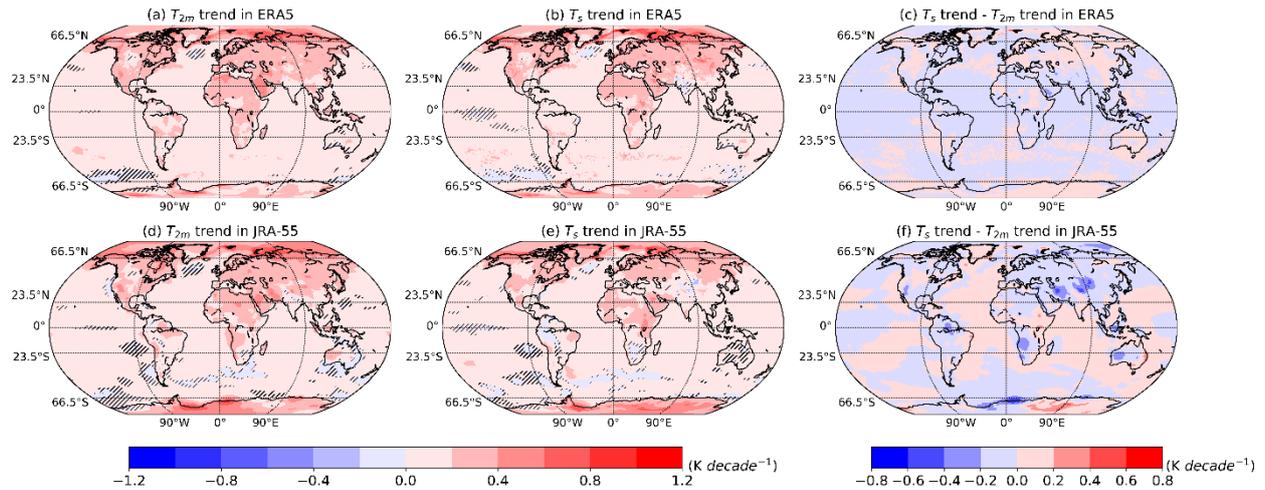

**Figure 3.** Trends of surface air temperature ($T_{2m}$) (first column) and surface skin temperature ($T_s$) (second column) from 1958 to 2021 for ERA5 **(a, d)** and JRA-55 **(b, e)**. The differential trends between $T_s$ and $T_{2m}$ trends for ERA5 **(c)** and JRA-55 **(f).** The hatch areas indicate trends that are not statistically significant. Trend units: K dec$^{-1}$.

Temperature trends (for $T_s$ and $T_{2m}$) over different latitude bins in both datasets agree with each other when viewing the spatial distribution of magnitudes globally (Fig. 4), manifested by much larger warming in the Arctic at a rate (all temperature rates given as K dec$^{-1}$) of ~0.55, which is more than three times warming than the global average rate (~0.15) (Figs. 4m, 4p). During the 1979-2021 period, this Arctic amplification of warming was markedly pronounced, occurring at rates 3.5 times faster than the global average warming rate documented by surface climate observations (Rantanen et al., 2022) in ERA5 data and 4 times faster in JRA-55 data. The potential causes of the Arctic amplification may be linked to sea ice decline, changes in atmospheric and oceanic heat contents, or changes in atmospheric moisture transport (Graversen, 2006; Screen and Simmonds, 2010; Zhang et al., 2008). Antarctic warming is at a rate of 0.2 K dec$^{-1}$ over the past 64 years but it should be noted that this warming rate is greater than 0.6 after 1980, which is consistent with surface observations (Rantanen et al., 2022). The datasets exhibit a clear consensus regarding the increased temperature of both land and ocean areas over the years. The land experienced greater warming at a rate exceeding 0.2 for both $T_s$ and $T_{2m}$ compared to the oceans' warming, leading to a land-to-ocean warming ratio ranging from 1.8 to 2.3 (Fig. 4). The explained reasons for the land-ocean contrast in warming rates is the influence of temperature and humidity on lapse rate, which results in a smaller reduction in lapse rate over land areas compared to the ocean, leading to a greater warming effect on land surfaces than on ocean surfaces (Joshi et al., 2008). Additionally, the restriction of moisture in the land boundary layer directly contributes to an intensified warming of land surfaces, which, in turn, raises the lapse rate over the land (Joshi et al., 2008).

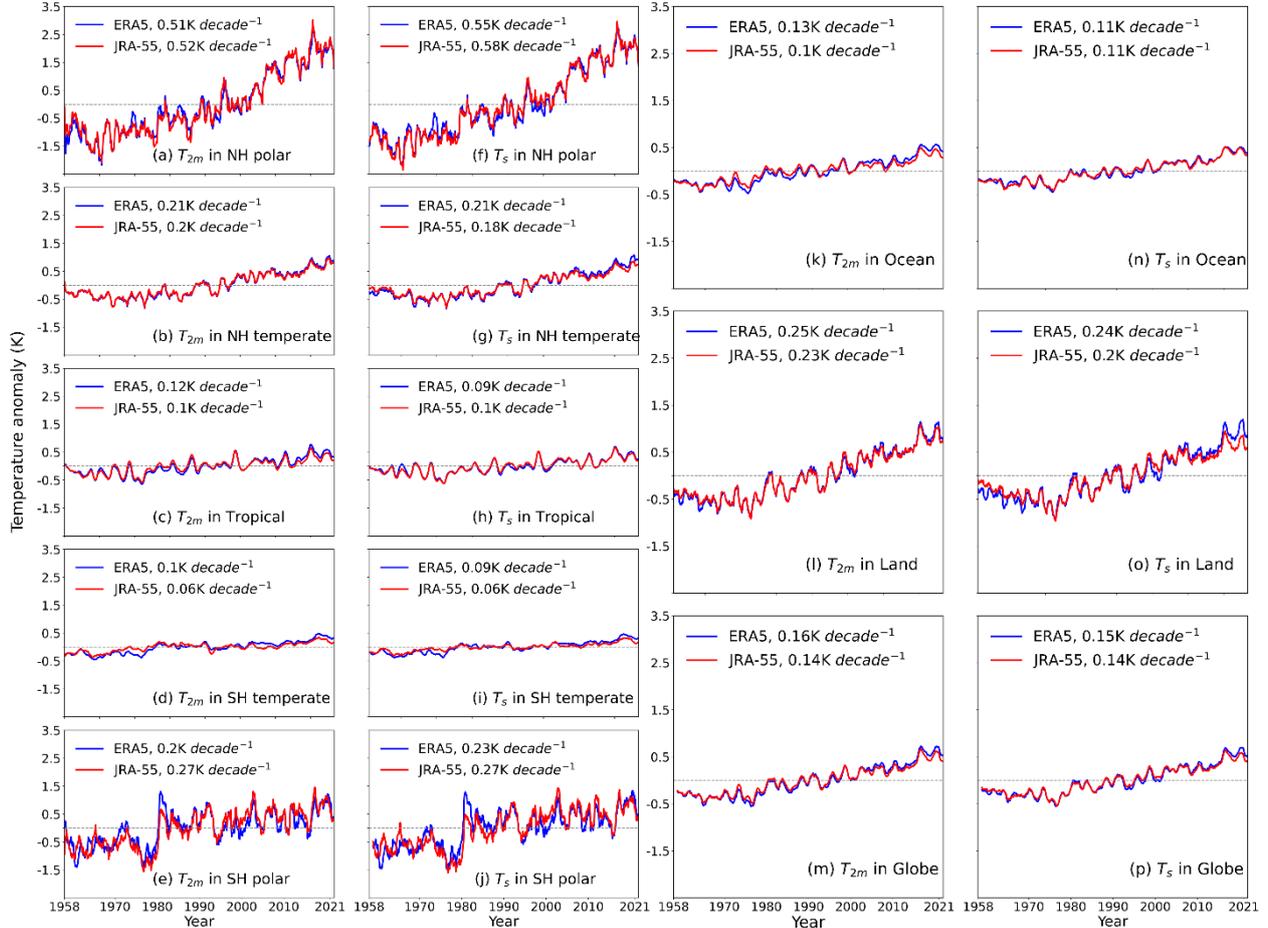

**Figure 4.** Monthly $T_{2m}$ and $T_s$ anomaly over **(a, f)** the Northern Hemisphere **(NH)** polar, **(b, g)** NH temperate, **(c, h)** Tropical, **(d, i)** Southern Hemisphere **(SH)** temperate, **(e, j)** SH polar, **(k, n)** ocean, **(l, o)** land, and **(m, p)** the globe from the ERA5 (blue curves) and JRA-55 (red curves) data sets during the period of 1958 to 2021. A 12-month running smoother was applied to all time series. All trends are statistically significant at a 95% confidence level.

## 3.3 TPW change response to temperature

TPW is expected to increase with air temperature by about 7% per K according to the Clausius–Clapeyron equation if the relative humidity in the lower troposphere is constant (Trenberth et al., 2005). This relationship is determined by the ratio (dTPW/dT) of TPW trends and temperature trends (Fig. 5). The dTPW/dT ratio shows a similar pattern between $T_{2m}$ and $T_s$ but there is a minor distinction between two reanalysis datasets. For ERA5, the TPW increases around 7.7%, 5%, and 6.5% per K for ocean, land, and globe, and close to 6% per K at a global scale based on observational HadCRUT5 (Allan et al., 2022), while these ratios are larger in JRA-55 (10.7%, 6.2%, and 8.6% per K, respectively). The dTPW/dT is generally around 6-8% per K over North America and NH polar regions for both ERA5 and JRA-55 (Fig. 5). However, TPW increase patterns with rising surface temperature do not always follow the Clausius-Clapeyron relation, especially over some areas of Asia and Europe where dTPW/dT ranges below 3 % per K and is even negative in central Africa (Fig. 5). Negative dTPW/dT occurs over the southern tropical ocean, where sea surface temperature was increased (Fig. 3) but precipitable water appears to be decreased slightly (Fig. 1).

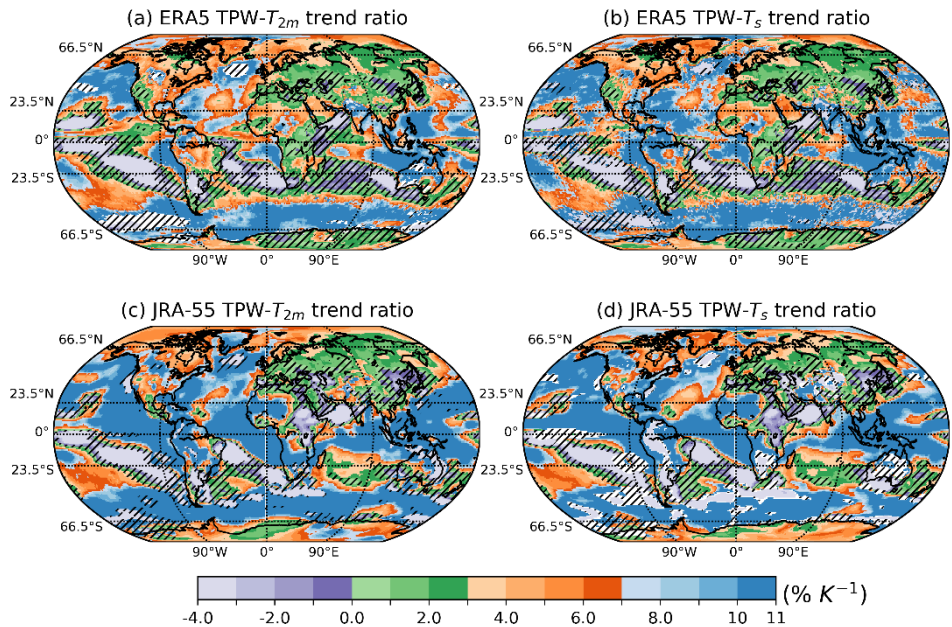

**Figure 5. (a)** The ratio (% $K^{-1}$) of the TPW trends and the surface air temperature ($T_{2m}$) trends from 1958 to 2021 in ERA5. **(b)** Same as **(a)** but for surface skin temperature (**$T_s$**). **(c)** Same as **(a)** but in JRA-55. **(d)** Same as **(a)** but for $T_s$ in JRA-55. The hatch areas represent ratios that are not statistically significant at the 95% confidence level.

In addition to the global patterns of spatial dTPW/dT ratios, the ratio's latitude dependency is an interesting subject (Fig. 6) between 60° S and 60° N. Although the TPW response values show strong variations across latitudes, the ratios show a similar changing pattern over the globe and ocean, which varies across the theoretical Clausius-Clapeyron curves of dTPW/dT ratios (Figs. 6a, c, d, f). Over most of the NH, both dTPW/dT ratios are smaller than the Clausius-Clapeyron response curves, but they are close to the Clausium-Clapeyron relationship over temperate latitudes (-40 to -20ºN band) in the SH for both the globe and ocean scales. Such strong latitude dependency and the discrepancy between land and ocean areas are associated with zonal relative humidity changes and possible amplification of surface warming over land relative to the ocean (O'Gorman and Muller, 2010) (Figs. 6c, f). There are two stronger TPW response zones, located in the southern high-latitude and the tropics over the globe and ocean (Figs. 6a, c, d, f). This result is nearly the same when drawn from multiple models of climate simulations for both historical and projected climate scenarios (O'Gorman and Muller, 2010). For land areas, in addition to two stronger response zones similar to the globe and ocean, a local maximum was found in the sub-tropical areas of the NH (Fig. 6). The ratios for the land region are mainly lower than the theoretical ratio (around 7% per K) and these ratios are even negative in tropical latitudes. Comparing these ratios based on $T_s$ and $T_{2m}$, their discrepancies are greater over land, which probably contributes to their discrepancies over the globe (Fig. 6).

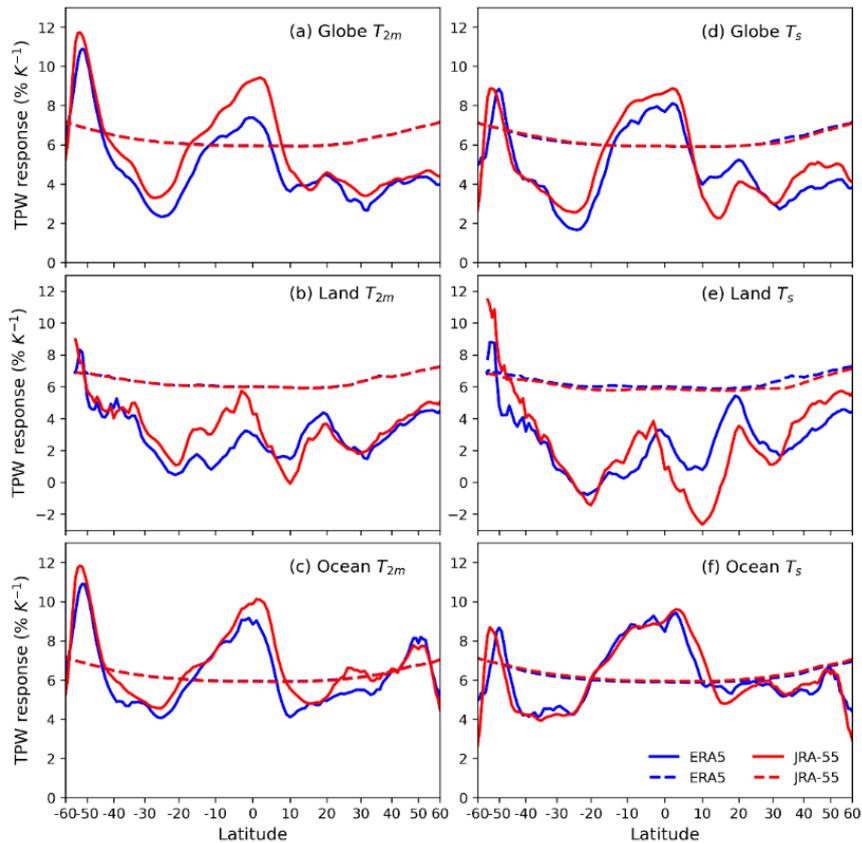

**Figure. 6**. The meridional means of change of TPW response to $T_{2m}$ (first column) and $T_s$ (second column) for ERA5 (blue lines) and JRA-55 (red lines) over **(a)** globe, **(b)** land, and **(c)** ocean. The dashed lines represent the theoretically expected Clausius-Clapeyron response based on the climatological zonal mean temperature from the trend analysis.

## 4 Further Discussion

The reanalysis dataset combines various observations to create a coherent global dataset using an atmospheric general circulation model. It provides spatially coherent and consistent data with global coverage, which makes it valuable for climate research (e.g., Allan et al., 2022; Urraca and Gobron, 2023). By assimilating multiple data sources and employing a physics-based model, reanalyses can effectively address uncertainties and minimize the presence of unrealistic values. However, it is worth mentioning that discontinuities in the time series may still occur in reanalysis datasets when satellites and conventional observations are transited, newly added, and removed during the assimilation process (Long et al., 2017). Therefore, we used the PMF method to test potential discontinuities of the reanalysis dataset over ocean, land, NH polar, North America, and tropical land regions where both TPW and temperature trends increased significantly (Figs. 1 and 3). These potential discontinuity points are summarized (see Tables S1-3) (Hersbach et al., 2020; Kobayashi et al., 2015). Testing results indicate no significant discontinuities of TPW over the ocean in JRA-55, NH polar, and tropical land regions in ERA5 (see details in Table S1). The maximum magnitude of discontinuity in ERA5 occurred in January 1977 over the ocean region, coinciding with the change in the assimilation data from DMSP14 SSMI to DMSP 15 SSMI. In JRA-55, the maximum discontinuity's magnitude was in July 2003, which matches with the assimilation of AMSU and MHS satellite data in June 2003 (Kobayashi et al., 2015; Uppala et al., 2005). Most of discontinuities in TPW occurred after the satellite era (i.e., after 1979) when massive satellite data were introduced. It seems that 38% of the TPW discontinuities can be explained by changes that occurred in surface observing systems but 62% of discontinuities are not matched with changes in data sources or satellite changes during assimilation in reanalysis datasets (see Table S1). Due to these undocumented discontinuities, it is not recommended to use adjusted or statistically homogenized timeseries for trend analysis without metadata confirmation (Wang and Feng, 2013) although there are slight differences between adjusted trends and raw trends (see Table S4).

For the surface temperatures, most of the discontinuities occur before the satellite era, therefore the long-term trend analysis for temperatures in reanalysis datasets after 1979 should be more reliable. In terms of temperature discontinuity identified (see Table S2-S3), the longer temperature trends should be more reliable by using surface observation networks' datasets (Pielke et al., 2007). Overall, $T_s$ over the NH polar region in JRA-55, TPW over the ocean in JRA-55, as well as TPW over NH polar and tropical land regions in ERA5, are homogenous. Discontinuities are detected in other regions that may result in minimal alterations to the trends. This leads us to infer that, although the adjustments do enhance consistency during certain specific periods and reduce disparities concerning reanalysis data at the monthly and yearly scales, the adjusted data's trends in our study could often cause inconsistency due to undocumented discontinuities.

The RHARM radiosonde observations (Madonna et al., 2022), which are mostly located in the NH and available from 1979-2019, are completely independent of reanalysis data. Compared with TPW trends from reanalysis data, moistening TPW in the NH polar region and Asia show the best agreement with radiosondes, with more than 90% and 60% of stations' trend differences within ± 0.2 mm dec$^{-1}$, respectively (Fig. S1). In addition, TPW trend differences are within ± 0.2 mm dec$^{-1}$ at more than 54% of stations in Europe and North America (Fig. S1). Moistening TPW trends in Western Europe and South Asia and drying TPW trends in Western North America are consistent with the results from both reanalysis data and GPS observations (Parracho et al., 2018). Overall, ERA5 and JRA-55 show good agreement with observations at regional scales. Specifically, TPW shows better agreement in North America and the NH polar region.

After evaluating the reanalysis data with radiosonde observations on land, water vapor data from AIRS and SSMI(S) satellites are used to evaluate the accuracy of reanalysis over oceanic regions from 2003 to 2021 (Fig. S2). Except for the drying trends of the Northern Atlantic Ocean in the short-term period, the spatial drying trend patterns in the two reanalysis datasets are similar to the long-term trends (Fig.1) although they are not statistically significant. The tropical oceans increased in water vapor at a rate of 1.6 % dec$^{-1}$ for ERA5 and 1.9 % dec$^{-1}$ for JRA-55 on average, with a similar moistening rate shown in SSMI(S) (1.1% dec$^{-1}$) but not in AIRS data (0.05% dec$^{-1}$). This result is consistent with Allan et al. (2022). The good agreement between the reanalyses and SSMI(S) is likely be attributed to the assimilation of all-sky radiances collected by the SSMI(S) satellite into the ERA5 and JRA-55 (Hersbach et al., 2020; Kobayashi et al., 2015). Although radiance measurement from AIRS is also introduced in ERA5 but not in JRA-55 (Hersbach et al., 2020; Kobayashi et al., 2015), the TPW from AIRS does not produce a strong global moistening and disagrees with reanalysis data and SSMI(S) microwave data and climate models, as well as Global Navigation System GPS (Allan et al., 2022; Douville et al., 2022).

## 5 Conclusions

Atmospheric reanalyses are widely used in the assessment of global climate change and their accuracy has advanced in recent years. Our study, bolstered by the latest ERA5 and JRA-55 reanalysis datasets, presents notable trends in the total precipitable water (TPW), surface temperature ($T_s$), and 2-meter temperature ($T_{2m}$) from 1958 to 2021. Beginning with regional TPW trends, we identified statistically significant moistening trends over North America, tropical land (±23.5° band), and Northern Hemisphere polar regions. These findings, which are in concert with prior studies (Borger et al. 2022; Trenberth et al. 2005; Wang et al. 2016), point to the solitary exception of the South Pacific Ocean, where TPW trends registered as negative. When considering TPW anomalies, we noted a consistent rise in both reanalysis datasets. Specifically, the moistening rates in ERA5 and JRA-55 were as follows: 0.57, 0.9, and 0.66 % per decade, and 0.84, 1.05, and 0.88 % per decade over ocean, land, and globe, respectively.

Building upon the analysis of TPW, our study further explored the temperature trends based on $T_{2m}$ and $T_s$ from ERA5 and JRA-55. Despite a general global warming trend, certain areas of the Southern Hemisphere and the North Atlantic Ocean demonstrated cooling. Terrestrial regions displayed a faster warming rate compared to oceanic regions, with a ratio of roughly 2:1, corroborating findings from Swaminathan et al. (2022). Arctic warming was particularly pronounced, registering three times the global average during 1958-2021, and escalating to around four times from 1979 to 2021. While Antarctic warming was more modest at 0.2 K per decade over the past 64 years, a sharp increase to over 0.6 K per decade was observed post-1980.

Last, we examined the TPW response to surface temperature changes, noting deviations from the Clausius-Clapeyron relation. In the ERA5 dataset, we identified TPW increases of around 7.7%, 5%, and 6.5% per K for ocean, land, and globe, respectively. These increased rates were higher in the JRA-55 dataset, with values at 10.7%, 6.2%, and 8.6% per K, respectively. Importantly, these response ratios were not uniform globally, presenting a variation between 6 and 8% $K^{-1}$ between latitudes 15 - 60ºN and increasing toward southern high latitudes over oceans. The most substantial ratios for deviations in TPW responses were discovered in the southern high latitudes across land, ocean, and the globe.

*Code and data availability.* All data used for this study are freely available. The ERA5 dataset is openly available from ECMWF: https://cds.climate.copernicus.eu/cdsapp#!/dataset/reanalyses-era5-single-levels-monthly-means?tab=overview. The JRA-55 dataset is available from https://rda.ucar.edu/datasets/ds628.0/. The AIRS satellite data is from https://disc.gsfc.nasa.gov/datasets/AIRS3STD_006/summary. The SSMI(S) satellite is from www.remss.com/missions/ssmi. The Radiosounding HARMonization dataset is from https://cds.climate.copernicus.eu/cdsapp#!/dataset/insitu-observations-igra-baseline-network?tab=doc.

*Author contributions.* NW performed the analysis of the results and the visualization. XL was responsible for the funding acquisition. The original manuscript was written by NW and revised by XL, RPs, XZ, and AN.

*Competing interests.* The corresponding author has stated that there are no competing interests among the authors.

*Acknowledgments.* We thank Dallas Staley for editing and finalizing the paper.

*Financial support.* This study was supported in part by the U.S. Department of Agriculture, Agricultural Research Service (A22-0103-001) and National Science Foundation (grant no. FAIN:2345039). The contribution number of this manuscript is 23-057-J.